\documentstyle[12pt]{article}

\newcommand{\EQ}{\begin{equation}}
\newcommand{\EN}{\end{equation}}
\def\ov{\overline}

\def\ga{\gamma}

\def\la{\lambda}
\def\eps{\epsilon}
\def\to{\rightarrow}
\def\gev{\; {\rm GeV}}
\def\ev{\; {\rm eV}}
\def\mev{\; {\rm MeV}}

\def\dem{\delta m^2}
\def\anue {\bar{\nu}_e}
\def\anu {\bar{\nu}}
\def \prl {Phys. Rev. Lett. }
\def \plb {Phys. Lett. B }

\def \prd {Phys. Rev. D }

\def \nue{\nu_e}
\def \numu{\nu_\mu}
\def \nutau{\nu_\tau}
\def \en {E}
\def \et {E_{\rm th}}
\def \sat {\sin^2 2 \theta}

\newcommand{\bea}{\begin{eqnarray}}
\newcommand{\eea}{\end{eqnarray}}
\newcommand{\bean}{\begin{eqnarray*}}
\newcommand{\eean}{\end{eqnarray*}}

\begin{document}
\topmargin 0pt
\oddsidemargin=-0.4truecm
\evensidemargin=-0.4truecm
\renewcommand{\thefootnote}{\fnsymbol{footnote}}
\setcounter{page}{1}
\begin{titlepage}
\begin{flushright}
INFN FE-11-94 \\
hep-ph/9409464\\
September 1994
\end{flushright}
\vspace{0.7cm}
\begin{center}
{\Large Vacuum oscillation solution to the solar
neutrino problem \\
\vspace{0.2cm}
in standard and non-standard pictures}
\vspace{1.2cm}

{\Large Zurab G. Berezhiani} \footnote{E-mail: 31801::berezhiani,
 berezhiani@ferrara.infn.it}\\
\vspace{0.2cm}
{\em Istituto Nazionale di Fisica Nucleare, sezione di Ferrara,I-44100
Ferrara,
Italy\\

and\\

Institute of Physics, Georgian Academy of Sciences, Tbilisi
380077, Georgia}\\
\vspace{0.4cm}
and \\
\vspace{0.4cm}
{\Large Anna Rossi} \footnote{E-mail: 31801::rossi, rossi@ferrara.infn.it } \\
\vspace{0.2cm}
{\em
Istituto Nazionale di Fisica Nucleare, sezione di Ferrara, I-44100
Ferrara,Italy}\\

\end{center}
\begin{abstract}
The neutrino long wavelength (just-so) oscillation is revisited as a
solution to the solar neutrino problem. We consider just-so scenario
in various cases: in the framework of the solar models with relaxed
prediction of the boron neutrino flux, as well as in the presence of
the  non-standard weak range interactions between neutrino and matter
constituents. We show that the fit of the experimental data in the
just-so scenario is not very good for any reasonable value of the
$^8B$ neutrino flux, but it substantially improves if the non-standard
$\tau$-neutrino--electron interaction
is included. These new interactions could also remove the conflict
of the just-so picture with the shape of the SN 1987A neutrino spectrum.
Special attention is devoted to the potential of the future real-time
solar neutrino detectors as are Super-Kamiokande, SNO and BOREXINO,
which could provide the model independent tests for the just-so scenario.
In particular, these imply specific deformation of the original solar
neutrino energy spectra, and time variation of the intermediate energy
monochromatic neutrino ($^7Be$ and $pep$) signals.

\vspace{0.4cm}

\end{abstract}

\end{titlepage}
\renewcommand{\thefootnote}{\arabic{footnote}}
\setcounter{footnote}{0}
\newpage
{\bf \large 1. Introduction}

\vspace{0.5cm}

The deficit of the solar neutrinos, dubbed the Solar Neutrino Problem
(SNP), was observed more than 20 years ago in the Homestake $Cl-Ar$
experiment. The 1970-93 average of the chlorine experiment result reads
as \cite{Davis}
\EQ
\label{cl}
R_{Cl}=2.32 \pm 0.26   ~\mbox{SNU}
\EN
whereas the Standard Solar Model (SSM)
by Bahcall and Pinsonneault (BP) \cite{BP} implies $R_{Cl}= 8$ SNU,
where $6.2$ SNU comes from $^8B$ neutrinos,   1.2 SNU from $^7Be$
neutrinos and the remaining 0.6 SNU from the other sources.
The predictions of the other SSM \cite{B,TC,CDF} do not differ strongly.
However, the chlorine result alone does not seem sufficient to pose the
problem, since the predicted flux of the boron neutrinos
has rather large uncertainties. These are  mainly due to the poorly known
nuclear cross sections $\sigma_{17}, \sigma_{34}$ at low energies,
some other astrophysical uncertainties which could change the solar
central temperature,  the plasma effects etc. (see e.g. \cite{BB} and
refs. therein). All these, working coherently,  may decrease
$\phi^B$ by more than a factor 2  compared to the SSM prediction.
Also the $^7Be$ neutrino flux can have uncertainties up to 20 \%.
Therefore, for a comprehensive analysis, it is suggestive to consider
these fluxes as free parameters: $\phi^B=f_B\phi^B_0$, $\phi^{Be}=f_{Be}
\phi^{Be}_0$, where $\phi_0$ are the BP model fluxes  and the
factors $f$ reflect the uncertainties.

However, the direct observation of solar $^8B$
neutrinos by Kamiokande detector  \cite{Hirata} brings another
evidence to the SNP. The Kamiokande signal  is less than
that is expected from the SSM by BP,  unless $f_B\leq 0.6$.
However, more important is that the signal/prediction ratio
\EQ
\label{k}
Z_K=\frac{R^{exp}_K} {R^{pred}_K} =
\frac{1}{f_B}\,(0.51\pm 0.07)
\EN
for any $f_B$ is incompatible to the one of the chlorine experiment
\EQ
\label{ZCl}
Z_{Cl}= \frac{R^{exp}_{Cl}} {R^{pred}_{Cl}} =
\frac{1}{0.78 f_B+0.22 f_{Be}}\, (0.29\pm 0.03)
\EN
unless $f_{Be}\ll f_{B}$ (for the simplicity, we have extended
the factor $f_{Be}$ also to other sources contributing the $Cl-Ar$
signal).  However, such a situation is absolutely improbable from
the astrophysical viewpoint: whatever
effect (e.g. diminishing the central temperature)
kills $^7Be$ neutrinos, it should kill more the $^8B$ ones.\footnote{
In fact, $f_{Be}/f_{B}$ could be diminished down to 0.75 if there exists
a very low energy resonance in the  $^3He+^3He$  cross section
\cite{res}. This, however, cannot reconcile the solar neutrino data.}

One could even assume that the uncalibrated Homestake experiment
has some uncontrollable systematical error and the true value of
$\phi^B$ is measured by Kamiokande (i.e. $f_B\approx 0.5$).
However, the  data of the $Ga-Ge$ experiment show that in doing so
the SNP will not disappear. Indeed, the weighted average of the
GALLEX \cite{Gallex} and SAGE \cite{Sage} results is:
\EQ
\label{ga}
R_{Ga}= 78\pm 10 ~\mbox{SNU}
\EN
as compared with the BP prediction $131$ SNU.
The bulk of this signal ($71$ SNU) comes from the $pp$ source.
The latter is essentially determined by the solar luminosity and,
therefore, cannot be seriously altered by astrophysical uncertainties.
On the other hand, the contribution of about $7$ SNU is
granted by the $^8B$ neutrinos as measured by Kamiokande.
Therefore, there is not much  room left for the  $^7Be$ neutrinos
which, according to BP model, have to provide $36$ SNU: $\phi^{Be}$
should be suppressed much stronger than $\phi^B$ ($f_{Be}< 0.25$).
Thus, the SNP which arised initially  as the boron neutrino problem,
now has become the problem of the beryllium neutrinos.

All these arguments are strong enough to believe that the
astrophysical solutions to the SNP are excluded \cite{BB}.
It is more conceivable that in the way to the earth the solar $\nue$'s
are partially converted into the other neutrino flavours.
Moreover, the experimental data require the conversion mechanism
capable to suppress differently neutrinos of different energies.
According to a  general paradigm, following from the
experimental results, it should lead to  a moderate
reduction of the $pp$ and $^8$B neutrino fluxes and to
a strong depletion of the intermediate energy $^7$Be flux.

The neutrino oscillation picture can provide the necessary energy
dependence in two regimes, which are known as the MSW \cite{MSW} and
the just-so \cite{GP,BPW1} scenarios.\footnote{
According to a {\it clich\'e}, the neutrino oscillation is
regarded as a non-standard property. However, from the
viewpoint of the modern particle physics, the
existence of the neutrino mass and mixing should be considered
as a rather standard feature. In the framework of the Standard Model
(SM) the neutrino mass can arise through the higher order operators of the
type $\frac{1}{M} (lCl)HH$, where $l$ and $H$ are respectively the
lepton and Higgs doublets and $M$ is some regulator scale.
In particular, the neutrino mass range needed for the just-so scenario
corresponds to the Planck scale, $M\sim 10^{19}\gev$, whereas
the  MSW scenario requires $M$ to be of  the order of
the supersymmetric grand unification scale, $M\sim 10^{16}\gev$.
As for the adjective "non-standard", it
should be rather reserved for  the really non-standard neutrino
properties, implied by the SNP solutions based on the
magnetic moment transition \cite{VVO} or on the fast
neutrino decay \cite{BCY}.}
The MSW resonant conversion in matter is the
most attractive and elegant solution, requiring $\dem$
of about $10^{-5}\ev^2$ and small mixing angle,  $\sat\sim 10^{-2}$.
It provides a very good fit of the experimental data, due to the
selective strong reduction of the $^7Be$ neutrinos \cite{L,mauro}.

Another attractive possibility is offered by
the just-so oscillation, i.e. vacuum oscillation $\nue\to\nu_x$
($\nu_x=\numu, \nutau$) with the wavelength comparable to the sun-earth
distance \cite{GP,BPW1}. This solution needs $\dem$ of about
$10^{-10} \ev^2$ and large mixing angles \cite{BPW2},
which parameter range can  be naturally generated
by non-perturbative quantum gravitational effects \cite{GB,ABS}.
The just-so  scenario, due to the energy dependence of the
survival probability,  can provide an acceptable fit of
the solar neutrino data (not as good, however, as the MSW does).
The recent analysis of this scenario is given in refs. \cite{KP}.

As it was pointed out in ref. \cite{smirnov},
this scenario faces the difficulty being  confronted
with the SN 1987A neutrino burst \cite{sn}.
The original $\anu_{\mu,\tau}$ energy spectrum from the
supernova has a larger average energy (about $25\mev$)
than the spectrum for $\anue$ (about $12\mev$), due to the smaller
opacities of $\anu_{\mu,\tau}$. The neutrino conversion $\anue\to \anu_x$
induced by the neutrino mixing results in a partial permutation of the
original $\anue$ and $\anu_x$ spectra. If the permutation
is strong, it would
significantly alter the energy spectrum of the supernova $\anue$-signal.
The analysis \cite{smirnov}, derived by using the SN 1987A data and
different models of the neutrino burst, shows that for $\dem\sim
10^{-10}-10^{-11}\ev^2$ the range of mixing excluded
at 99\% CL is $\sat\geq 0.7$,
which covers the range required by the just-so scenario, $\sat\geq 0.7$.
Neverteless, we do not consider the SN 1987A argument as a
sharp evidence against the large neutrino mixing.
Moreover, as we will discuss below, this constraint can be removed
by assuming some non-standard neutrino interactions which could
increase the $\anu_x$ opacity in the supernova core, reducing
thereby its average energy.

In the present  paper we address certain  issues
in the context of the
long wavelength neutrino oscillation as a possible solution to the SNP.
In Sect. 2 we study how  this scenario fits the
experimental data in various cases:
(i) SSM+SM: in the reference SSM by Bahcall and Pinsonneault \cite{BP},
(ii) NSSM+SM: in the context of models with relaxed
prediction of $\phi^B$ (which we conventionally refer to as non-standard
solar models). In both cases the neutrinos are
supposed to have only the standard interactions,
(iii) SSM+NSM: in the SSM framework, assuming however that neutrinos have
some additional non-standard interactions with matter constituents.

Sect. 3 is devoted to the model independent analysis of the just-so
scenario. This essentially implies the modification of the solar
neutrino spectrum due to the energy and time dependence of the
survival probability.
We focus our attention on the advantages inherent in the future
real-time neutrino detectors like
Super-Kamiokande \cite{SK}, SNO \cite{SNO} and BOREXINO \cite{Borex}.
All these experiments can measure the recoil electron spectrum,
which could provide specific signatures
allowing to discriminate the just-so scenario, in particular from
the MSW one.

At the end, we give a brief summary of our conclusions.

\vspace{0.6cm}

{\bf 2. Data fit in standard and non-standard pictures}

\vspace{0.4cm}

For the simplicity, we consider the vacuum oscillations in the
case of two neutrino flavours: $\nue\to \nu_x$, where $\nu_x$ can be
$\numu$ or $\nutau$.\footnote{Certainly, the general case of three neutrino
oscillations involves more parameters. However, in many interesting cases
the three neutrino oscillation picture effectively reduces to the case of
two neutrinos. For example, in the case of the democratic ansatz of the
gravitationally induced neutrino mass matrix \cite{ABS}, the oscillation
picture is equivalent to the case of two neutrinos with $\sat= 8/9$.}
The survival probability  for solar $\nue$'s with energy $\en$ is given by:
\EQ
P(L_t,E)= 1-\sat \sin^2(\pi\,\frac{L_t}{l})
\label{surp}
\EN
where $l=\frac{4\pi E}{\dem}=
\frac{E[\mev]} {\dem[10^{-10}\ev ^2]}\cdot 2.47\cdot 10^{10}\,\rm{m}$
is the oscillation wavelength. The sun-earth distance $L$ depends on time as
$L_t= \ov{L}[1 -\varepsilon\cos (2\pi t/T)]$, where
$\ov{L}=1.5 \cdot 10^{11}$m, $T=365$ days, and
$\varepsilon=0.0167$ is the ellipticity of the orbit.

The time averaged signals predicted in the radiochemical experiments
is given by:
\EQ
\label{rate1}
R= \int dE \sigma (E) \sum_i \langle P(E) \phi^i \rangle _{T}\la_i(E)
\EN
Here $\sigma(E)$ is the detection cross section, $\phi^i$ are
the fluxes of the relevant components of the solar
neutrinos ($i=B, Be,$ etc.), $\la_i(E)$ are their
energy spectra normalized to 1,  and $\langle \dots \rangle_T$ stands
for the average over the whole time period $T$.
In this way, the time dependence of the original flux
($\phi(t)\propto L^{-2}_t$) is also taken into the account.

For the Kamiokande detector, since we consider the $\nue$ conversion
into an active neutrino, the expression for the signal becomes
\EQ
R_K=  \int_{\et} dE \la_B(E) \left[ \langle P(E)  \phi^B\rangle _{T}
 \sigma_{\nue}(E) +\left( \langle \phi^B \rangle _{T}
- \langle P(E)  \phi^B\rangle _{T}\right)
\sigma_{\nu_x}(E) \right]
\label{rate2}
\EN
Here $\sigma_{\nu_y}$ $(y=e,x)$ is the $\nu_y e^-$ scattering cross
section and
$\et= \frac{1}{2}( T_e+ \sqrt{T_e(T_e +2m_e)})$, where $T_e=7.5\mev$
is the recoil electron kinetic energy threshold.

Below we examine the just-so scenario in view of the recent status of
the solar neutrino problem. We accept the hypothesis that the solar
neutrino luminocities are constant in time, and
use the averaged data of the chlorine, gallium and Kamiokande
experiments to perform the standard $\chi^2$ analysis for various cases
(for the run-by-run analysis see ref. \cite{KP}.)

{\bf (i) SSM+SM.}
We use as reference SSM the BP model,
without taking into account the underlying theoretical uncertainties.
The case of the other SSM will be effectively recovered by relaxing
$\phi^B$ and $\phi^{Be}$ (see below).

The fit is not so  good:
the minimal $\chi^2$ obtained is 4.4. Thus, the just-so oscillation
is allowed as a SNP solution at the 3.6\% confidence level.
Once this solution is assumed, the parameter regions  containing
the true values with the
68\% and 95\% probability are given by
$\chi^2 \leq \chi^2_{min} + \delta \chi^2$, where
$ \delta \chi^2= 2.28,~ 5.99$ respectively.
These regions are shown in Fig. 1. They are limited by the  values
$\dem= (5-8)\cdot 10^{-11} \ev ^2$, $\sat= 0.7-1$. Our results
are essentially in agreement with the recent analysis \cite{KP},
where a somewhat different way of the data fitting is used.

In the same figure, we have also shown the $\dem$ dependence
of the time averaged $\nue\to\nu_x$ transition probability
for the monochromatic $^7Be$ and $pep$ neutrinos.
For the best fit point these probabilities are large, in agreement
with the general paradigm implying a strong  suppression for the
intermediate energy neutrinos. However, as we see, in the wide range
of the CL parameter regions there is no definite behaviour and even the
ratio of the signals (which can be measured in BOREXINO
detector -- see below) is unpredictable.
On the other hand,  the same
effect of the strong oscillation leads to the significant time
variations of these monochromatic neutrino lines (see below).


{\bf (ii)  NSSM + SM.}
Here $\phi^B$ and, to a less extent, also $\phi^{Be}$ are considered as
free parameters. So, we describe the $^8B$ neutrino flux as
$\phi^B= f_B\cdot \phi^B_{0}$,
where evidently $\phi_{0}^B$ is the prediction of the BP SSM
and the factor $f_B$ accounting
for the uncertainty  is varied in the range $0.4-1.6$
(for example, by taking $f_B=0.8$, $f_{Be}=0.9$ the case of
the Turk-Chi$\grave{e}$ze and Lopez SSM \cite{TC} is reproduced).
The lower limit $f_B=0.4$ is actually set  by
the Kamiokande measurement of the boron neutrino flux.

We have repeated the $\chi^2$ analysis for varyous values of $f_B$ and
$f_{Be}$. The corresponding best fit points and 68\% CL parameter areas
are given in Fig. 2.
The relevant range of $\dem$ remains rather stable against  variation
of $f_B$, whereas the $\sat$ becomes smaller with decreasing
$f_B$.\footnote{The analysis of ref. \cite{KS} shows that
the MSW scenario reacts in the same way by varying $f_B$, but the
best $\chi^2$ fit is achieved for $f_B=1$.}
The lowering (increasing) of $f_B$ results in a weakening
(strengthening) of the neutrino oscillations.  Therefore, with
smaller values of $f_B$ the model could be in agreement with the
SN 1987A bound $\sat\leq 0.7$ \cite{smirnov}.
However, as a general tendency, by decreasing $f_B$ the fit becomes
worse, whereas it slightly improves for $f_B>1$.
E.g., for $f_B=0.4$ the high value of $\chi^2_{min}=11.7$ indicates a
poor fit (this solution is excluded at more than 99.9\% CL).\footnote{
It is interesting to note that for $f_B\simeq 0.6$ even the one
parameter  ($\sat$) fit of the averaged short-wavelength oscillation
provides somewhat better CL, $\chi^2_{min}=11$ at $\sat=1$. }
On the contrary, for $f_B=1.3$ we have $\chi^2_{min}= 3.0$ which is
acceptable at 8.3\% CL. In this case the boron neutrino flux must
be depleted stronger so that the larger mixing
is required, what reconciles mutually the chlorine
and the Kamiokande data.  On the other hand, the large mixing
contradicts the supernova bound.
The decreasing of the beryllium flux (see Fig. 2b) does not alter
significantly the previous results.

{\bf (iii) SSM+NSM.}
Here we still take the BP model as reference SSM but assume that
neutrinos have some non-standard interactions in addition to the SM
ones. Namely, we suppose that the $\nu_x$ state in which the solar
$\nue$ is converted  is just $\nutau$ and it has extra weak range
interaction with the electron:\footnote{For $\numu$
such interactions are severely restricted by laboratory limits
(see \cite{sci} and refs. therein.) }
\EQ
\label{nsm}
{\cal L}_{eff}=-\frac {G_F}{\sqrt{2}}\,
\anu _{\tau} \ga^\mu (1-\ga_5)\nutau \left [\eps\,
\bar{e}\ga_\mu  (1+\ga_5) e +
\eps' \,\bar{e}\ga_\mu (1-\ga_5) e  \right]
\EN
Here $\eps$ and $\eps'$ parametrize the strength of new
interactions with respect to the Fermi constant $G_F$.
The first term in this lagrangian, with positive $\eps$,
can be effectively obtained
(after the Fierz transformation) from the exchange of some
additional electroweak doublet scalar $\varphi$ (the
relevant Yukawa coupling is $\bar{l}_{\tau L} e_R \varphi$, where
$l_{\tau L}$ is the lepton doublet including $\tau$ and $e_R$ is the right
handed component of the electron). The second term
could be due to the exchange of some charged singlet Higgs $\eta$.
However, the same exchange of the charged singlet unavoidably
contributes the $\tau\to e\nu_{\tau}\bar{\nu}_e$ decay
width, which sets the strong bound $\eps'< 0.05$.
As for the strength of the first interaction $\eps$, its value
is not seriously constrained by any laboratory limit, while
the astrophysical bounds on stellar evolution
in the most conservative case imply $\eps\leq 1$ \cite{sci}.

The  extra neutral current
interaction  of $\nutau$ with the electron contributes
to the $\nutau-e$ elastic scattering together with the
standard neutral current and, as far as $\eps>0$, it increases
the $\sigma_{\nutau}$ cross section (see below, Fig. 9), and
thus the signal in the Kamiokande detector.
This implies a larger suppression of
the boron neutrino flux, what leads to a better agreement between
the Kamiokande and Homestake data.

In order to study the impact of these extra NC coupling on
the just-so scenario,\footnote{The effects of such non-standard
interactions (flavour diagonal as well as flavour changing) for
the MSW picture were studied in ref. \cite{FC}. However,
the altering of the neutrino
propagation in the solar interior due to the interaction (\ref{nsm})
has no importance in the case of just-so oscillation. This interaction
is relevant only for the detection cross-section in the
$\nu-e$ scattering experiment.}
we have repeated the $\chi^2$ analysis for the interval $\eps=0-1$.
The results of the fitting are shown in Fig. 3.
One can observe that the allowed region of the  parameters
$\dem$ and $\sat$ is rather stable against the variation of $\eps$.
However, as it was expected the data fit
improves by increasing $\eps$, since  now the Kamiokande signal requires
larger mixing angles. E.g., for $\eps=1$ we achieve
$\chi^2_{min}= 1.8$, which implies that in this case the just-so oscillations
can be regarded as a solution of the SNP at the 18\% CL.

Certainly, along with the interactions (\ref{nsm}) one can consider
also the analogous non-standard interactions of $\nutau$
with protons and neutrons. They could be induced
due to the exchange of some scalar leptoquark with mass
of about 100 GeV. These interactions do not
contribute the signal in the detectors under operation.
Nevertheless, they can be relevant for the signal in the
future real-time detectors, expecially SNO and BOREX.

Let us conclude this section with the following remarks.
As we have seen, the just-so picture can be relevant for SNP only
for the following mass and mixing range
\EQ
\dem=(0.5-0.8)\cdot10^{-10}\ev^2\,,~~~~\sat=0.7-1
\label{interval}
\EN
for any reasonable values $f_{B,Be}$ and $\eps$
(see Figs. 2,3). Moreover, for the plausible interval $f_B=0.7-1.3$
the best fit area is essentially located in the very narrow band
around $\dem\approx 0.6\cdot 10^{-10}$ eV$^2$, rather independently
on the concrete values of $f_{B,Be}$ and $\eps$, while $\sat$
varies from 0.7 to 1 depending on the concrete values of these
parameters. The data fit for certain  cases of the simultaneous
variation of $f_B$ and $\eps$ is shown in the Table 1.

\newpage

{\bf 3. Predictions for the future solar neutrino experiments }

\vspace{0.4cm}

Although the data fit in the just-so scenario is somewhat worse than
in the MSW picture, it cannot be ruled out as a SNP solution.
On  the other hand, these  solutions cannot be discriminated by
the recent experiments. However, the next generation of
the solar neutrino detectors will shed more light on the situation.
The novel detectors like Super-Kamiokande \cite{SK}, SNO \cite{SNO}
and BOREXINO/BOREX \cite{Borex} could provide tests, almost
independent of the SSM details. In particular, these real time detectors
will be able to observe the seasonal time variations of the various neutrino
components, due to the ellipticity of the earth orbit and
sufficiently strong (but not {\em very} strong to be averaged)
oscillation effects in the just-so regime. On the contrary,
the MSW mechanism can exhibit only the standard 7\% {\em simultaneous}
variation of all signals from  June to December, since in this case
all neutrino conversions take place in the sun interior and the
small oscillation effects in the  way from sun to earth are negligible.

As we have seen, the just-so picture can
be relevant for SNP only for a narrow interval (\ref{interval}),
 rather independently on the values $f_{B,Be}$ and $\eps$
(see Figs. 2,3). Moreover, for the moderate values $f_B=0.7-1.3$
the best fit area is essentially located at
$\dem\approx 0.6\cdot 10^{-10}$ eV$^2$.
Then it is easy to see that for the $\dem$ in the range
(\ref{interval}) the monochromatic $^7Be$ neutrinos ($E=0.861$ MeV)
oscillate along the distance $\ov{L}=1.5\cdot 10^{11}$ m about $3-5$
times, $pep$ neutrinos ($E=1.442$ MeV) about $2-3$ times and the boron
neutrinos (with typical energy of about 10 MeV) do not undergo even
one full oscillation. Therefore, since the value $\varepsilon\pi\ov{L}/l$
is a small parameter (e.g., for $^7Be$ neutrinos it is about 0.2),
from eq. (\ref{surp}) we obtain for the $\nu_e$ survival probabilities
at June and December ($L_{\pm}=\ov{L}(1\pm \varepsilon)$):
\EQ
P_{\pm}(E) \approx \ov{P}(E)\mp [1-\ov{P}(E)]\,
\frac{2\varepsilon\pi\ov{L}/l_E}{\tan (\pi\ov{L}/l_E)}
\label{varia}
\EN
where the quantity $\ov{P}(E)=P(\ov{L},E)$ essentially is the average
survival probability of the $\nu_e$ with energy $E$.
This formula demonstrates that the seasonal variations should be
stronger for the neutrinos with smaller energies, and it
can be dramatic for the monochromatic neutrino lines \cite{GP,BPW1}.
Namely, in the best fit region ($\dem\approx 0.6\cdot 10^{-10}$ eV$^2$)
we have $\ov{P}_{Be}\sim 0.5$ for the $^7Be$ neutrinos (see Fig. 1)
while the phase factor  $\tan (\pi\ov{L}/l_{Be})\sim 1$. Therefore,
one should expect up to 50\% seasonal variations for the beryllium
neutrino signal in BOREXINO detector (see below, Table 1).
The standard 7\% variations are negligible in this case. At the same time,
for this range of $\dem$ the variation of the $pep$ signal is
expected to be smaller, less then 10\%, essentially due to large
$\tan (\pi\ov{L}/l_{pep})$ -- see Fig. 1. However, for
the wider range of parameters (\ref{interval}) also the $pep$
neutrino signal variation can be significant. As for the $^8B$ neutrinos,
one cannot expect strong time variations (up to 10\%), due to
large oscillation length as well as smoothing effects due to
continuous spectrum.\footnote{The feasibility of the Super-Kamiokande
and SNO detectors for the observation of the boron neutrino signal
variations was studied in details in the recent paper
by Krastev and Petcov \cite{Plamen}. }

Another possibility to discriminate the just-so scenario is
related to the spectral distortion of  the various solar neutrino
components. The original  energy spectra  $\la_i(E)$ ($i=B,Be,$ etc.)
are independent of the details of the solar models. They are determined
only by the nuclear reactions producing the neutrinos. The neutrino
energy dependent conversion mechanisms for the SNP solution
can strongly modify the initial neutrino spectra,
offering thereby specific signatures for their discrimination.
Below we consider the ``just-so" spectral predictions
for the planned experiments.

{\bf Super-Kamiokande.} This detector is expected to measure the spectrum
of the high energy $^8B$ neutrinos.
The original neutrino distribution can be reproduced from
the recoil electron spectrum due to $\nu- e$ scattering, though it
is somewhat smeared  due to the integration over the neutrino energy:
\bea
 F (T)\!\! = \!\! \int  dE \la_B(E)
\left[ \langle P(E) \phi^B\rangle _{T}\frac{d\sigma _{\nue}}{dT}(E,T)+
\left(\langle\phi^B\rangle _{T} - \langle P(E)\phi^B\rangle _{T} \right)
\frac{d\sigma_{\nu_x}}{dT}(E,T)  \right]
\nonumber  \\
\frac{d\sigma _{\nu_y}}{dT}(E,T)\!=\!  \frac{ 2G^2_F m_e}{\pi}~
\left[ g^2_{yL}  + g^2_{yR}\left( 1-T/E \right)^2 -
g_{yL}g_{yR}~ m_e ~T/ E^2 \right], ~~~y=e,\mu,\tau
\label {desigma}
\eea
where $T$ is the recoil electron energy.
For the $\nue-e$ scattering we adopt the Standard Model
values for the NC coupling constants $g_{eL}= \frac{1}{2}  +
\sin^2 \theta_W$ and $g_{eR}= \sin^2 \theta_W$, whereas for
the $\nu_x$ state we  also account for the possible non-standard
couplings given by eq. (\ref{nsm}):
$g_{xL}= -\frac{1}{2}  + \sin^2 \theta_W +\eps'$ and
$g_{xR}=  \sin^2 \theta_W+\eps$.

We calculate the ratio of the distorted spectrum $F(T)$ to that
is predicted by the SSM $F_0(T)$. For the definiteness we
normalize $\xi(T)=F(T)/F_0(T)$ to 1 at $T=10\mev$. Clearly,
this ratio does not depend on the  SSM details, as far as $F_0(T)$
is essentially determined by the boron beta decay spectrum $\la_B(E)$.

The shape of $\xi(T)$ for various couples of the parameters
$\dem$ and $\sat$ from the allowed area
is given in Fig. 4a for $\eps=0$ and Fig. 4b for $\eps=1$.
The present sensitivity of Kamiokande (long error bars)
is not enough to discriminate the just-so solution, whereas
Super-Kamiokande (short error bars) could distinguish it from the
MSW picture, expecially due to the
characteristic distortion in the lower energy part of the spectrum
(for the recoil electron spectrum in MSW case see ref. \cite{KS}).
The deformation of the energy spectrum can alter
the average energy  $\ov{T}$ of the recoil electrons as compared to the
standard spectrum prediction  $\ov{T}_0=7.44\mev$
(with an electron energy threshold $T_{th}=5.5\mev$).
In Fig. 5 the iso-curves for the variation of  $\ov{T}$ as compared
to $\ov{T}_0$ (in percents) are plotted in the $(\dem,\sat)$ plane.
As we see, $\ov{T}$ can change up to 4\%. In the case $\eps=0$
(Fig. 5a) the variation is rather positive than negative, whereas
for $\eps=1$ (Fig. 5b) it is dominantly negative.
In particular, for the best fit solutions  the variation is
$2.6$ \% for $\eps=0$,  and $-0.6$\% for $\eps=1$.

{\bf SNO.}
This heavy water real-time detector will measure the $^8B$ neutrino flux
through the charged-current (CC) and neutral-current (NC) processes:
\bea
& & \mbox{CC}:~~~~ ~ \nue d \to e^- p~ p \label{cc} \\ \nonumber
& &  \mbox{NC}:~~~~~  \nu_y d \to \nu_y p~ n ~~~,~~ y=e,\mu,\tau
\eea
The ratio $\eta=R_{CC}/R_{NC}$ in the SSM (i.e. when
no neutrino conversion takes place)  is independent
of the value of $f_B$. If the neutrino conversion occurs, the flux of
the survived solar $\nue$ is directly measured by the CC signal:
\EQ
R_{CC}= \int_{\et} dE \sigma_{CC} (E) \la_B(E)
\langle P(E) \phi_B \rangle_T
\label {scc}
\EN
where $\et=7\mev$ and for the cross section $\sigma_{CC}$
we  use the data presented in \cite{KN}.

If the solar $\nue$'s are converted into active neutrinos
$\nu_x=\numu,\nutau$ having only the SM neutral current couplings
to nucleons ($Z$-boson exchange), then the probability conservation
guarantees that the NC signal is the
same as in the reference SSM:  $R_{NC}$ directly
measures the original $\phi^B$ flux. Therefore, if the measured ratio
$\eta=R_{CC}/R_{NC}$ is less than that is predicted by SSM
($\eta_0=1.8$ for $\et= 7\mev$, independently of $f_B$), this
would unambiguously indicate the deficit of the boron $\nue$, caused
by the neutrino conversion.
In Fig. 7 the iso-signal curves
are given for the ratio $Z_{SNO}=R_{CC}/R^{pred}_{CC}= \eta/\eta_0$.
As we see, in the parameter region relevant for the just-so scenario
this ratio varies in the range $0.2-0.35$.

The CC signal will allow to clearly discriminate the just-so picture
by measuring the recoil electron spectrum $F(T)$. In fact, the latter
reproduces the  energy spectrum of the $\nu_e$'s survived the conversion,
i.e. $\langle P(E)\phi_B \rangle_T \la_B(E)$, shifted by an amount equal
to the small recoil energy left to the nuclei: $T=E-1.44\mev$. Therefore,
the ratio of the distorted spectrum to the SSM predicted one
does not depend on $f_B$ and it directly characterizes the energy
dependence of the survival probability.

In Fig. 6 the ratio $\xi(E)=F(E)/F_0(E)$,
normalized to 1 at $E=10\mev$, is plotted for the same
parameters as in the Fig. 4. The presence of the pronounced minimum
discriminates the just-so solution from the MSW one, which instead
provides characteristic monotonic shape of this ratio \cite{KS}.
The effect is manifested stronger than in the case of Super-Kamiokande,
since now the spectral distortion is not smoothed by the integration
over the neutrino energy. In Fig. 7 we show the iso-curves of
the recoil electron average energy deviation from the SSM prediction
($\ov{T}_0=8.42\mev$ with the electron energy threshold of $5.5\mev$).
It ranges up to 12\%, stronger than in Super-Kamiokande.
For the best fit points it is $\sim8\%$ for $\eps=0$ and $11.5\%$ for
$\eps=1$ (see Table 1).
The energy variation in the MSW picture
has the same sign  \cite{mauro}, but it is considerably smaller.

The non-standard interactions (\ref{nsm}) of $\nu_\tau$ with electrons
do not contribute the signal neither in CC nor NC channels.
However, the presence of the analogous non-standard $\nu_\tau$ interactions
with quarks,
violating universality of the neutrino interactions with nucleons, could
be relevant. In this case the neutral current signal becomes
\EQ
R_{NC}= R^{SM}_{NC}+ \int_{\et} dE \Delta\sigma^{NSM}_{NC} (E)
\la_B(E) \left(
\langle \phi^B \rangle_T-\langle P(E)\phi^B\rangle_T
\right)
\label {snc}
\EN
where $\Delta\sigma^{NSM}_{NC}$ is the additional (to the SM)
contribution to the $\nu_x d \to \nu_x p~ n$ cross
section arising due to the non-standard interactions.
This extra contribution can differently affect the ratio
$\eta$ expected, depending on the sign of $\Delta\sigma^{NSM}_{NC}$.
In particular, in the case of sterile $\nu_x$ (i.e. when the extra
contribution exactly cancels the standard one), we have $\eta \approx
\eta_0$ independently of whether the conversion occurs or not
\cite{Carlo}.

{\bf BOREXINO.} Due to the high radiopurity of this scintillator, the
detection threshold is low: $T=0.25\mev$.
This allows to have enough statistics to detect the $^7Be$
and $pep$ neutrino lines  through the $\nu-e$ scattering. In fact, the
beryllium neutrino flux can be measured by exploring the energy
window $T=0.25-0.7\mev$ for the recoil electrons. In this window,
according to BP SSM,  about 50 events are expected per day, versus
about 10 events provided by the natural radioactivity background
 \cite{Borex}. As for the $pep$ neutrinos whose contribution dominates
the recoil electron energy range $T=0.7-1.3\mev$, their detection is
less feasible, since the predicted signal (about 3 events per day)
is comparable with the internal background.

As already anticipated, in the just-so picture the strong
oscillations of the intermediate energy neutrinos prevent to make some
definite prediction for the time averaged signals of $^7Be$ and $pep$
neutrinos: in the relevant parameter region the signal to the SSM
prediction ratios $Z_{Be}$ and $Z_{pep}$ can be rather arbitrary
(see Fig. 1). In the MSW case, no precise prediction can be obtained
as well \cite{mauro},
however, the relation between the $^7Be$ and $pep$ signals remains
close to that is expected in SSM.\footnote{As we commented above,
the best possibility to distinguish MSW and just-so scenarios
is provided by strong seasonal variations of $^7Be$ and $pep$ neutrino
signals in the later case. }
 On the contrary, in  the case of just-so
solution no definite prediction can be given neither for these signal
ratio $Z_{Be}/Z_{pep}=\frac{R_{Be}/R_{pep}}{[R_{Be}/R_{pep}]_0}$:
in the relevant
parameter regions it can be much less or more than 1 (see Fig. 1).

The high sensitivity of the BOREXINO detector will  allow
to measure the recoil electron energy spectrum due to the $^7Be$
neutrinos and, to some extent, also due to the $pep$ ones.
In this respect it is of interest to study how these spectra are
affected in the just-so oscillation picture. The typical curves of the
$\nu-e$ event distribution for some  parameter values are plotted in
Fig. 8a,b for the cases $\eps=0$ and $\eps=1$. In the former case,
when $\nu_\tau$ has only SM interactions with the electron, the energy
spectrum appears generally depleted throughout the relevant energy interval.
However, the shape of the spectrum is not substantially changed and it
essentially repeats the one of SSM (see Fig. 8a).
In the case of NSM the rate of events is less depleted
in the $^7Be$ energy window: in the presence of new interactions the
$\nutau$ contribution becomes very effective for the lower energies,
which compensates the deficit of the original $\nue$'s.
Moreover, for $\eps\simeq 1$ the signal  can be even larger than that is
expected in SSM: $Z_{Be}>1$ (see Fig. 8b). Also, the shape of the
spectrum becomes steeper as compared to the SSM predicted one.
Let us remark also that the compensating effects of the $\tau-$neutrino
NSM interactions can smear the time variations of $^7Be$ and $pep$
neutrino signals (compare the Tables 1A and 1B).


\vspace{0.5cm}

{\bf 4. Discussion }

\vspace{0.5cm}

We have confronted the just-so oscillation scenario with the recent
experimental data on the solar neutrinos experiments in the context
of non-standard solar models. Namely, we studied the response of this
scenario to possible changes of the boron and beryllium neutrino
fluxes.
In the framework of the BP SSM the data fit is not excellent:
$\chi^2_{min}=4.4$, while it becomes worse for $f_B<1$ and slightly
improves for $f_B>1$. The better data fit can be achieved by assuming
that the $\nu_x$ state, emerged from the oscillation, has some non-standard
neutral current coupling to the electron.
The existing laboratory and astrophysical bounds
indeed allow the $\tau-$neutrino to have such NSM interactions
in the weak range, with $\eps \simeq 1$. In this case, also with
moderate increasing of $f_B$ (up to 1.3), one can achieve quite
reasonable $\chi^2$ fit (see Table 1B).
It is interesting to note that
the relevant mass range is rather stable against the variation of
$f_{B,Be}$ and $\eps$: for the best fit area we have
$\dem \approx 6\cdot 10^{-11}\ev^2$.

The new generation of the real-time solar neutrino detectors can
test the just-so scenario independently of the SSM details,
and distinguish it from other candidates to the  SNP solution.
Even more, the possible NSM neutrino interactions can be also tested.
Indeed, these detectors will be able to measure the spectra
of various solar neutrino components, as well as to detect the effects
of their seasonal variations. This will allow to determine
unambiguously all unknown parameters, namely the SSM ones
($f_{B,Be}$ etc.), possible NSM ones ($\eps$, etc.) as well as
neutrino mass and mixing range itself. In  Table 1 we show the
average rates
and their seasonal variations in the chlorine, gallium
and  Kamiokande experiments,
as well as in the future detectors (Super-Kamiokande,
SNO and BOREXINO), for the best fit points corresponding to different
values  of $f_B$ and $\eps$.

In the case of $\nue\to \nu_x$ just-so oscillation the recoil electron
energy spectra appear to be specifically altered, and
different from the one expected from the MSW conversion. Let us
imagine that the SNO and/or Super-Kamiokande spectral measurements
really point to the just-so oscillation. These spectra separately cannot
tell us anything about the presence of the NSM interactions
of $\nu_x$ with the electron (compare the curves in Figs. 4a and 4b).
However, both the CC and NC reactions in SNO provide
the measurements of the boron neutrino energy spectrum on the earth,
which also constitutes the only contribution to the Super-Kamiokande
signal.
Therefore, the presence of non-standard $\nu_x-e$ interactions could be
determined by confronting the spectra measured by SNO and
Super-Kamiokande. In fact, the CC reaction in SNO directly measures the
energy spectrum of the survived boron $\nue$'s reaching the earth, i.e.
essentially the value $P(E) \phi_B$. Substituting this value in the
eq. (\ref{desigma}) for the Super-Kamiokande signal, one can
unambiguously deduce the only `unknown' quantity, the  differential
cross-section $\frac{d\sigma_{\nu_x}}{dT}(E,T)$, and confront it to
the SM prediction. (As we mentioned above, the
non-standard $\nu_x-e$ couplings can be also tested by the
spectral shape of the recoil electrons in BOREXINO.)
By confronting the CC and NC signals in an analogous
manner, one can extract also the information on possible NSM
couplings of $\nu_x$ with nucleons (see eq. (\ref{snc})).
Thus, as far as we believe that SNP is related to some conversion mechanism
of solar $\nue$'s into the other neutrino flavours, the sun appears to
be quite a strong and cheap source of the latter.
Then the measurement of the recoil electron energy spectra in the novel
real-time detectors offers not only a test
for any possible SNP solution, but it can also be considered as a test
for the neutrino NSM interactions, or in other words, as a test for
the Standard Model of the electroweak interactions itself.

Last but not least we wish to emphasize that the non-standard neutrino
interactions, besides improving the data fit in the just-so picture,
could also resolve its potential conflict with the SN 1987A $\nu$-signal,
pointed out in ref. \cite{smirnov}.
Namely, these interactions would increase the $\anu_\tau$ opacity in
the supernova core, and thereby reduce their average energy.
This could occur due to the dramatic increase of the $\anu_\tau-e$
cross-section, as compared with the $\nu_\tau-e$ one,
for large values of $\eps$ (see Fig. 9, where these cross-sections
are plotted versus the neutrino energy for different values of $\eps$).
Then the interference  of the original $\anue$ and
$\anu_\tau$ spectra due to the neutrino mixing will less affect
the expected $\anue$ signal in the terrestrial detectors. According
to ref. \cite{smirnov}, the problem will be dissolved
if the average energy of $\anu_\tau$ drops below $17-20\mev$: then even
the maximal mixing, $\sat=1$, cannot be excluded. Moreover, in this case
the partial permutation between the $\anue$ and $\anu_\tau$ spectra could
explain the certain excess of the higher energy $\anue$ events from
SN 1987A following from the comparison of the IMB and Kamiokande data
\cite{sn}.
On the other hand, the difference between $\anu_\tau$ and
$\nu_\tau$ opacities can provide a significant asymmetry in their
average energies, which, due to the strong oscillation, can result in
an asymmetry between $\anue$ (isotropic) and $\nue$ (directional)
signals in the terrestrial detectors.
Obviously, for the precise evaluation of the effects from
the non-standard neutrino interactions it is necessary to
 consistently  include
them into a detailed computer analysis of the stellar core collapse
at the beginning.

\vspace{0.7cm}

{\bf Acknowledgements.}

\vspace{3mm}

We are grateful to G. Fiorentini for illuminating  conversations.
Useful discussions with
S. Degl'Innocenti, G. Di Domenico, S. Petcov, B. Saitta
and A. Smirnov  are also gratefully acknowledged.
When our paper was under preparation, we received the preprints by
Krastev and Petcov \cite{Plamen}, and Bilenky and Giunti \cite{Giunti},
devoted to the same subject. Our analysis differs from
these ones in many aspects. In particular, we studied the case of
non-standard solar models as well as the impact of the possible
non-standard neutrino interactions.
We thank C. Giunti and P. Krastev for sending their preprints
and useful comments.

\newpage


\vspace{-0.5cm}
{\small
\begin{table}[hbt]
\begin{center}

\begin{tabular}{|l|c|c|c|c|c|c|} \hline
{}~~~~{\large\bf A:} $\eps=0$  &\multicolumn{2}{c|} {$f_B=1~
(\chi^2_{min}=4.4)$} &
\multicolumn{2}{c|}{$f_B=0.7~ (\chi^2_{min}=6.4)$} &
\multicolumn{2}{c|}{$f_B=1.3~ (\chi^2_{min}=3.0)$}  \\
\cline{2-7}
 & $Z$ & $R$ & $Z$ & $R$ & $Z$& $R$ \\
\hline
$Cl-Ar$ & 0.32 ($\mp$ 10\%)& 2.55 & 0.43 ($\mp4.7$\%) & 2.63 & 0.27
($\mp 12$\%)& 2.65\\
\hline
$Ga-Ge$ & 0.50 ($\mp$10\%) & 66 & 0.55 ($\mp 4.3$\%)& 70 & 0.51 ($\mp 10$\%)
& 69\\
\hline
Kamiokande & 0.41 ($\mp 2.4$\%) & 0.41 & 0.52 ($\mp 2.0$\%)& 0.36
& 0.34 ($\mp 3.5$\%)& 0.44\\
 ($T_{th}=7.5\mev$)& [0.31+0.1]& & [0.44+0.08]& &(0.23+0.11] & \\
\hline
SK & 0.37 ($\mp$ 2.0\%) & 0.37 & 0.49 ($\mp1.1$\%) & 0.34
& 0.28 ($\mp 3.2$\%)
& 0.36\\
 ($T_{th}=5.5\mev$)& [0.26+0.11]& & [0.40+0.09]& &[0.16+0.12] & \\
\hline
SNO & 0.26 ($\mp3.7$\%) & 0.26 & 0.41 ($\mp 1.5$\%)& 0.29 & 0.17
($\mp 6.5$\%)& 0.22\\
\hline
BOREXINO: $^7Be$ & 0.61 ($\mp$22\%) & 32 & 0.55 ($\pm$11\%) & 29 &
0.79 ($\pm$ 18\%)& 41\\
($T=0.25-0.7\mev$)& [0.45+0.16] & & [0.39+0.16]&
&[0.67+0.12] & \\
\hline
BOREXINO: $pep$ & 0.41 ($\pm4.7$\%)& 1.0 & 0.58 ($\pm 6.2$\%)& 1.5 &
0.32 ($\mp 3.2$\%)& 0.8\\
($T=0.7-1.3\mev$)& [0.28+0.13]& & [0.48+0.10]& &[0.16+0.16] & \\
\hline
$\delta(T)_{SK}$ &\multicolumn{2}{c|} {$2.6\%$}  &
\multicolumn{2}{c|} {$1.4\%$} & \multicolumn{2}{c|} {$4.0\%$}\\
\hline
$\delta(T)_{SNO}$ &\multicolumn{2}{c|} {$8.0\%$}  &
\multicolumn{2}{c|} {$3.8\%$} & \multicolumn{2}{c|} {$14.4\%$}\\
\hline
\end{tabular}
\label{tab1a}
\end{center}
\end{table}

\vspace{-1.0cm}

\begin{table}[hbt]
\begin{center}

\begin{tabular}{|l|c|c|c|c|c|c|} \hline
{}~~~~{\large\bf B:} $\eps=1$  &\multicolumn{2}{c|} {$f_B=1~
(\chi^2_{min}=1.8)$} &
\multicolumn{2}{c|}{$f_B=0.7~ (\chi^2_{min}=4.2)$} &
\multicolumn{2}{c|}{$f_B=1.3~ (\chi^2_{min}=1.0)$}  \\
\cline{2-7}
 & $Z$ & $R$ & $Z$ & $R$ & $Z$& $R$ \\
\hline
$Cl-Ar$ & 0.31 ($\mp 13$\%) & 2.47 & 0.41 ($\mp5.5$\%)
& 2.51 & 0.25 ($\pm 8.0$\%)& 2.45\\
\hline
$Ga-Ge$ & 0.54 ($\mp 10$\%)& 71 & 0.55 ($\mp7$\%)& 70
& 0.55 ($\pm 10$\%)& 75\\
\hline
Kamiokande & 0.44 ($\mp2.0$\%)& 0.44 & 0.56 ($\mp1.3$\%)& 0.39 &
0.35 ($\mp 2.8$\%)& 0.46\\
 ($T_{th}=7.5\mev$)& [0.26+0.18]& & [0.42+0.14]& &[0.14+0.21] & \\
\hline
SK & 0.46 ($\mp 1.3$\%) & 0.46 & 0.58 ($\mp 0.7$\%)& 0.41 &
0.41 ($\mp 0.2$\%) & 0.53\\
 ($T_{th}=5.5\mev$)& [0.20+0.26]& & [0.37+0.21]& &[0.11+0.30] & \\
\hline
SNO & 0.21 ($\mp4.5$\%)& 0.21 & 0.38 ($\mp 1.5$\%)& 0.27 &
0.11 ($\mp 5.4$\%)& 0.14\\
\hline
BOREXINO: $^7Be$ & 1.02 ($\pm 2.0$\%)& 53 & 1.04 ($\pm$1.0\%)
& 54 & 1.02  ($\mp$1.5\%) & 53\\
($T=0.25-0.7\mev$)& [0.68+0.34]& &[0.42+0.62]&
&[0.80+0.21] &\\
\hline
BOREXINO: $pep$ & 0.72 ($\mp0.7$\%)& 1.8 & 0.80 ($\pm$1.3\%)& 2.1 &
0.92 ($\pm$ 2.7\%)& 2.4\\
($T=0.7-1.3\mev$)& [0.20+0.52]& & [0.42+0.38]& &[0.77+0.15]
& \\
\hline
$\delta(T)_{SK}$ &\multicolumn{2}{c|} {$-0.6\%$}  &
\multicolumn{2}{c|} {$-0.6\%$} & \multicolumn{2}{c|} {$-2.6\%$}\\
\hline
$\delta(T)_{SNO}$ &\multicolumn{2}{c|} {$11.5\%$}  &
\multicolumn{2}{c|} {$4.5\%$} & \multicolumn{2}{c|} {$13.0\%$}\\
\hline
\end{tabular}
\label{tab1b}
\end{center}
\end{table}

\vspace{-0.4cm}

{\bf Table 1.} The expected signals in different detectors,
for the best fit points corresponding to different values  of $f_B$.
The Tables {\bf A,B} are for the  cases  $\eps=0,1$, respectively.
$Z$ is the ratio of the calculated signal
to the one expected in the solar model with the given $f_B$ (clearly, $Z$
does not depend on $f_B$). Within round brackets the
percentage seasonal variation of the signal, compared to the time
averaged value
$Z$, is reported, where the upper sign refers to June and the lower one to
December.
For the $\nu-e$ scattering experiments
the individual contributions from the survived $\nue$ and emerged $\nu_x$
are also shown (within the square brackets).
$R$ are the annual average signals predicted for each
detector. For the radiochemical experiments $R$
is given in SNU, whereas for BOREXINO in the number of events per day,
for the recoil electron energy intervals indicated.
For (Super) Kamiokande and SNO $R$ is given in units of the BP SSM
prediction:  $R=f_B\cdot Z$.
The quantity $\delta (T)$ stands for the variations of the
recoil electron average energy with respect to the one predicted in SSM.

\newpage
{\bf Figure Captions}

\vspace{0.5cm}
{\bf Fig. 1.}
Confidence regions in the parameter space $\dem$ and $\sat$,
for the case SSM+SM. The diamond marks the best fit point
to the experimental data ($\chi^2_{min}=4.4$).
Solid and dotted curves delimit the
68 \% CL and 95 \% CL regions, respectively.
On the right axis, the time averaged transition probabilities
(modulo $\sat$) are shown as a function of $\dem$ for the $^7Be$
and $pep$  neutrinos (dashed  and dot-dashed curves, respectively).

\vspace {0.2cm}

{\bf Fig. 2.}
The best fit points (diamonds) and the 68 \% CL regions in the
case NSSM+SM for different $f_B$, where $f_{Be}=1$ (Fig. 2a)
or $f_{Be}=0.8$ (Fig. 2b).
The $\chi^2_{min}$ corresponding to values $f_B=0.4,~ 0.7,~1.0, ~
1.3,~ 1.6$ are $11.7, ~ 6.4, ~ 4.4, ~ 3.0, ~2.8$ in Fig. 2a, and
$10.3, ~ 5.7,~ 4.3,~ 3.1, ~2.8$ in Fig. 2b.

\vspace{0.2cm}

{\bf Fig. 3.}
The best fit point (marked as 2, $\chi^2_{min}=1.8$)
and the 68 \% CL regions in
the case SSM+NSM, for $\eps=1$ (solid curves) confronted with the case
SSM+SM, $\eps=0$ (dotted curves, best fit point marked as 1).
 In the following these points, as well as the other
typical points 3 and 4, will be used to demonstrate the effects of
spectral distortion.

\vspace{0.2cm}

{\bf Fig. 4.} Super-Kamiokande: the ratio $\xi(T)$ of the recoil
electron energy spectrum, distorted due to the just-so oscillation, to the
undistorted one (normalized to 1 at $10\mev$),
given for the points shown in Fig. 3.
%
Here Fig. 4a,b refer to the cases $\eps=0$ and $\eps=1$, respectively.
The longer error bars indicate the present sensitivity of the Kamiokande
detector and the shorter ones represent the expected sensitivity in
Super-Kamiokande.

\vspace{0.2cm}

{\bf Fig. 5.} The iso-signal curves for $Z_{SK}$
expected at Super-Kamiokande, with 5.5$\mev$ threshold (solid).
The curves for the iso-percentage variations  of the
average electron energy compared with the SSM value are also shown
(dashed).
Fig. 5a refers to the case $\eps=0$ and Fig. 5b to that  $\eps=1$.
The corresponding 68\% CL regions  are also shown (dotted curves).

\vspace{0.2cm}

{\bf Fig. 6.} SNO:  the ratio $\xi(E)$
of the distorted boron neutrino energy spectrum
to that expected in absence of solar neutrino
conversion, normalized to 1 at $10\mev$.
The  curves correspond to the points marked in Fig. 3.
The error bars indicate the expected sensitivity of the detector.

\vspace{0.2cm}

{\bf Fig. 7.} The iso-signal contours due to the CC reaction at SNO
with 5.5$\mev$ threshold (solid).
The dashed curves represent the iso-percentage variations
of the average electron energy as compared to that expected in SSM.

\vspace{0.2cm}

{\bf Fig. 8.} Distribution of the $\nu-e$ scattering events
expected at BOREXINO as a function of the recoil electron energy $T$,
for the cases $\eps=0$ (Fig. 8a) and $\eps=1$ (Fig. 8b). These are
given for the typical points shown in Fig. 3 (solid, long dashed,
dot-dashed and short dashed curves, respectively).
For comparison, the dotted curve corresponds the electron
spectrum expected in BP SSM, in the absence of neutrino conversion.

\vspace{0.2cm}

{\bf Fig. 9.} The energy dependence of the $\anu_\tau-e$
and $\nu_\tau-e$ scattering cross-sections (dashed and solid,
respectively), normalized to $\sigma_0= 2G^2_F m^2_e/\pi$,
for different values of $\eps$.

\end{document}